\documentclass{openjournal}
\usepackage{xcolor}
\usepackage{textgreek}
\usepackage[utf8]{inputenc}
\usepackage[english]{babel}

\usepackage{hyperref}
\hypersetup{
    unicode, 
    colorlinks=true,
    linkcolor=linkcolor,
    citecolor=linkcolor,
    filecolor=linkcolor,
    urlcolor=linkcolor,
}
\usepackage{color,colortbl}
\definecolor{linkcolor}{rgb}{0.0,0.3,0.5}
\usepackage{tensind}
\tensordelimiter{?}
\DeclareGraphicsExtensions{.bmp,.png,.jpg,.pdf}
\usepackage{verbatim}
\usepackage[normalem]{ulem}
\usepackage{orcidlink}
\usepackage{soul}
\usepackage{wrapfig}
\graphicspath{ {./figs/} }
\begin{document}
\title{Distinguishing cosmologies with topological analysis of 21-cm hydrogen 2D brightness fields at Cosmic Dawn}
\author{Maksym Tsizh\orcidlink{0000-0002-4480-3185}}
\email{maksym.tsizh@lnu.edu.ua}
\affiliation{Astronomical Observatory of Ivan Franko National University of Lviv, Kyryla i Methodia str. 8, Lviv, 79005, Ukraine}
\author{Anton Rudakovskyi\orcidlink{0000-0001-6161-3386}}
\affiliation{Dipartimento di Fisica e Astronomia, Università di Bologna,Via Gobetti 92/3, 40121, Bologna, Italy}
\affiliation{Bogolyubov Institute for Theoretical Physics, Metrolohichna 14-b, UA-02000 Kyiv, Ukraine}
\author{Vitalii Tymchyshyn\orcidlink{0000-0003-2292-600X}}
\affiliation{Bogolyubov Institute for Theoretical Physics, Metrolohichna 14-b, UA-02000 Kyiv, Ukraine}
\affiliation{Data Science Center, Kyiv Academic University, UA-03142 Kyiv, Ukraine}

\begin{abstract}
The 21-cm hydrogen line brightness field is a potentially valuable source of information about astrophysical parameters during the cosmological Dark Ages and the Epoch of Reionization. However, it can also be a source of observational data for cosmologists seeking to constrain cosmological parameters. While much of the effort has been focused on constraining parameters using information across a broad frequency range, most often with the power spectrum of inhomogeneities, we investigate whether a single frequency channel, corresponding to a single redshift slice, can distinguish between cosmological models. We utilize topological data analysis to evaluate the possibility of such constraints. In particular, we use cubical complexes to perform persistent homology analysis of the 21-cm line brightness field. We generate 21-cm brightness fields of different cosmologies with the 21cmFAST package and then select snapshots at several redshifts corresponding to the Epoch of Reionization. Our goal is to find the redshift that is most suitable for highlighting the diversity between scenarios within a certain range of cosmological parameters. To compare persistent pictures built on simulations of the brightness field, we use Wasserstein distance, computed between any two persistent diagrams. We find that observations of the 21-cm field near $z=10$ have the potential to distinguish between cosmological models with different values of the normalized Hubble parameter $h_0$, the spectral index of scalar matter perturbations $n_s$, the matter density parameter $\Omega_m$, and the matter fluctuation amplitude $\sigma_8$, even in the presence of instrumental thermal noise. We perform the PERMANOVA statistical test on the matrices of Wasserstein distances to confirm this hypothesis for all considered cosmological parameters.
\end{abstract}
\maketitle
\section{Introduction}
\label{sec:Intro}

The redshifted 21-cm line of neutral hydrogen is the most direct probe of the Cosmic Dawn and the Epoch of Reionization (EoR), the era in which the first luminous sources heated and ionized the intergalactic medium
\citep{furlanetto2006, pritchard2012}. Current and forthcoming low-frequency interferometers---LOFAR, MWA, HERA, and ultimately the Square Kilometer Array (SKA)---aim to measure the spatial fluctuations of the 21-cm brightness temperature across this epoch \citep{koopmans2015, deboer2017, mertens2020}. The standard summary statistic for these measurements is the power spectrum, which is statistically sufficient for a Gaussian random field but discards the substantial non-Gaussian information encoded in the morphology of the ionized
regions.

This motivates exploring summary statistics that are explicitly sensitive to the geometry and topology of the field. Morphological descriptors have a long history in cosmology, beginning with the genus of isodensity contours \citep{gott1986} and the Minkowski functionals \citep{mecke1994}, and extending to the Betti numbers \citep{park2013, pranav2017, tymchyshyn2024}, and, finally, to persistent homology \citep{pranav2019, tsizh2023, bermejo2024}. Topological data analysis (TDA), and persistent homology in particular, characterizes a manifold through the birth and death, across a filtration process, of its topological features, such as connected components, loops, and enclosed voids, providing a scale-free, parameter-less description of a complex structure of a given manifold. A persistence image of a field or pointcloud, with intermediary to a manifold, allows one to effectively reduce the amount of data, keeping the important connection and morphological information present that the power spectrum would probably omit.

In the broader cosmological context, TDA has emerged as a competitive and complementary summary statistic. Persistent homology of the large-scale structure constrains primordial non-Gaussianity and the standard cosmological parameters, in several cases outperforming the combined power spectrum and bispectrum and breaking parameter degeneracies \citep{biagetti2021, biagetti2022, yip2024}. Parameter inference directly from persistence images combined with machine learning has further improved these constraints while retaining interpretability, since each pixel of such an image corresponds to a definite region of the birth--persistence plane \citep{calles2024}. In weak gravitational lensing, it tightens constraints on $S_8$ and $\Omega_{\rm m}$ relative to the peak count method: cosmological constraints were obtained from the persistent homology applied to the cosmic shear survey data \citep{heydenreich2021, heydenreich2022}. Simulation-based inference based on persistent diagrams data shows a possibility to give tighter restrictions on $S_8$ and $\Omega_{\rm m}$ than those from cosmic shear two-point statistics \citep{prat2025}.

Several groups have also studied topological and morphological statistics of the 21-cm signal from the epoch of reionization. \citet{kapahtia2018, kapahtia2019, kapahtia2021} combined contour Minkowski tensors, Betti numbers, and the areas of excursion-set structures with a Bayesian analysis of 21cmFAST simulations. They showed that a mock SKA1-Low image at $z \approx 7.4$ ($\bar{x}_{\rm HI} \approx 0.5$) can constrain the ionizing efficiency, the minimum virial temperature of star-forming halos, and the maximum ionizing-photon mean free path.

The persistent topology of the reionization bubble network has been formalized and shown to add information beyond the power spectrum \citep{elbers2019, elbers2023}. In \citet{elbers2023}, the authors studied whether persistent homology of 21-cm observations can distinguish two reionization scenarios, one driven by faint galaxies and the other by bright galaxies. They found that tunnels ($\beta_1$ features) dominate the topology during reionization, and that their number is the easiest to observe and the most sensitive to the source properties. Their way of separating model differences from cosmic variance is similar to ours: they simulated several 21-cm lightcones of each model from different random initial conditions and compared the two groups with a permutation test, using the mean squared Wasserstein distance between persistence diagrams as the test statistic.

Betti numbers can track the characteristic percolation history of the ionized regions, as shown in \citep{giri2021}. Related morphological analyses based on shapefinders and percolation were performed in \citep{bag2018, bag2019}. Three-dimensional Minkowski functionals were used in \citep{diao2024} for reionization parameter inference. Minkowski functionals shapefinders were also utilized in \citep{ghara2024} to study the morphology of the binarized 21-cm field in heating-dominated Cosmic Dawn. In \citep{shimabukuro2025}, the authors analyzed a 1-dimensional continuous field of 21-cm forest, which is formed by X-ray heating from matter clusters, by counting 0-dimensional homologies obtained in a filtration process. However, all of the studies mentioned above vary astrophysical or small-scale dark-matter properties while keeping fixed the cosmological parameters that govern large-scale structure and the expansion history.

We, in turn, studied the 1-dimensional persistent homologies of 2-dimensional slices at a given redshift, corresponding to observations in individual frequency channels, similarly obtained with a filtration process, where filtration was based on the values of pixels of the field. In contrast to the authors of \citep{elbers2019, elbers2023}, who worked with a 3D 21-cm brightness field and studied the sensitivity of the method to the astrophysical reionization parameters, we concentrate on the dependence of persistent diagrams on cosmological parameters. We present our work with the following structure. In section \ref{cub} we explain the TDA approach and, in particular, the cubical complex filtration. Following it is section \ref{data} where we describe the data generated with 21cmfast, that is, our cosmological simulations. In section \ref{res} we describe the results. Section \ref{concl} is devoted to the discussion of the latter.

\section{Cubical Complex}
\label{cub}
Comparing scalar fields and spatial distributions is a fundamental challenge across the physical sciences.
A wide variety of comparison metrics have been developed for this purpose, including point-wise function norms such as the $L^p$ family, statistical distances between probability distributions such as the Kolmogorov--Smirnov statistic and the Kullback--Leibler (KL) divergence, geometric measures including the Fréchet distance, and transport-based metrics such as the Earth Mover's Distance (EMD). While these approaches are highly effective for many applications, they are generally not designed to characterize the multi-scale topological structure of the underlying data \cite{carlsson2009topology, ghrist2008barcodes, edelsbrunner2010computational}. However, as demonstrated by \cite{tsizh2023}, these multi-scale topological features can be sensitive to the underlying cosmological parameters. In the latter work, a discrete point distribution was transformed into a nested family of topological spaces indexed by a scale parameter $r$\,--- corresponding to the radius of balls centered at each point\,--- a process known as filtration. Topological invariants were computed for each space in the filtration, their evolution summarized as persistence diagrams, and compared using a Wasserstein distance specifically tailored for these diagrams in order to discriminate between cosmological models (for additional mathematical details, see the appendix of \cite{tymchyshyn2024}).  

The principal limitation of reusing that pipeline is its reliance on discrete point distributions rather than continuous scalar fields. Persistent homology on cubical complexes naturally overcomes this limitation by constructing a filtration from the sublevel sets of a scalar field sampled on a rectilinear grid \cite{carlsson2009topology,edelsbrunner2010computational}. This allows a continuous scalar field to be transformed into a nested family of topological spaces indexed by a threshold parameter (the level), while leaving the remainder of the pipeline intact. In our case (brightness fields), the level corresponds to the value of brightness temperature in each voxel of the simulations.

\begin{figure}[htbp]
    \centering
    \includegraphics[width=0.9\textwidth]{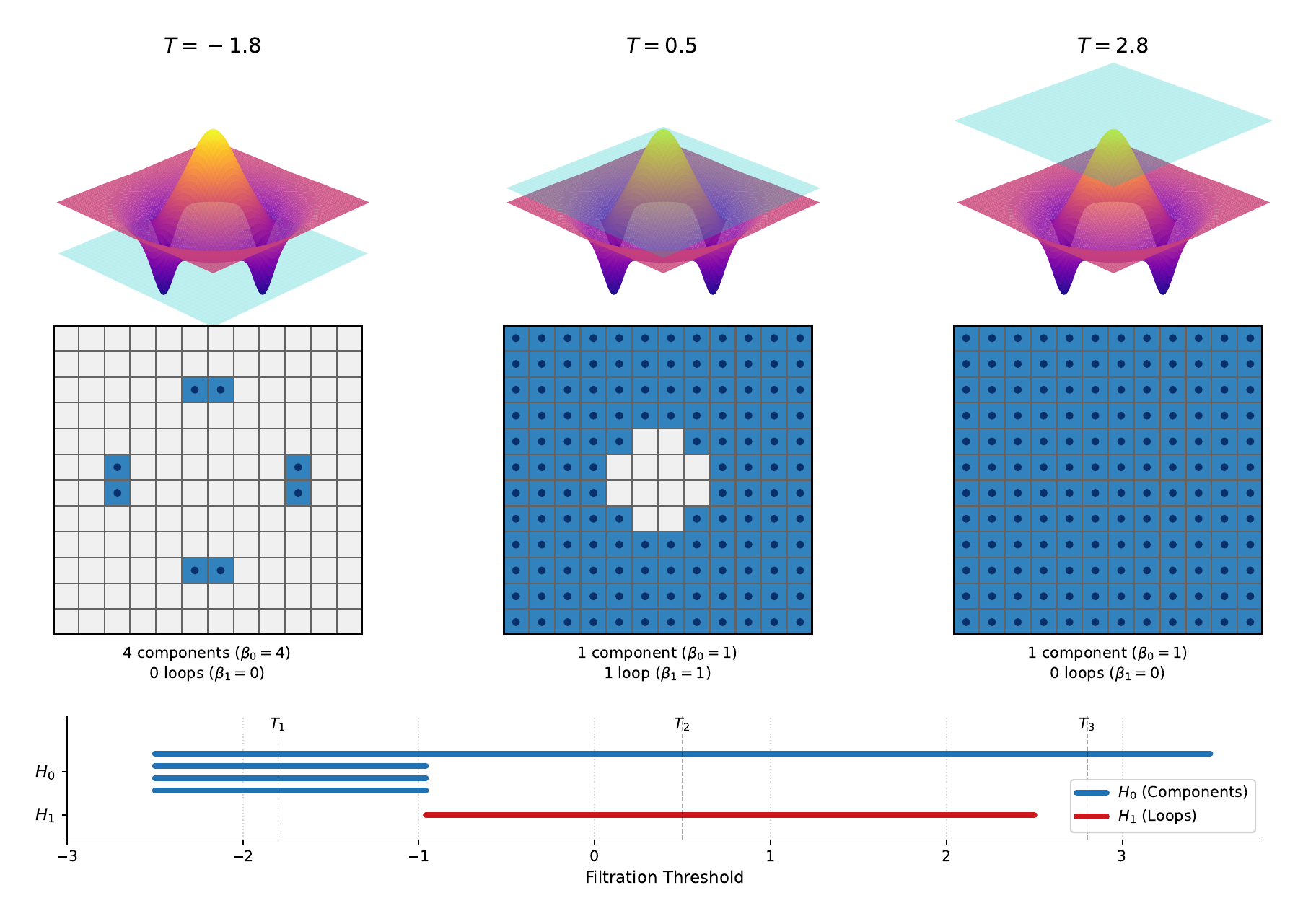}
    \caption{
    Illustration of a sublevel set filtration on a 2D scalar field mapped to a cubical complex. 
    Top row (3D): The scalar field is displayed with a cyan cutting plane representing the rising threshold level $T$. 
    Middle row (2D): The corresponding cubical complex; blue regions denote active cells below the threshold, and black grid lines illustrate the discrete lattice structure. 
    At Level 1 ($H_0$), four components exist. 
    By Level 2, they merge into a single component enclosing a large central void ($H_1$). 
    By Level 3, the void fills. 
    Bottom row: The resulting persistence barcode (blue bars: $H_0$, red bar: $H_1$). 
    Vertical dashed lines explicitly map the physical states in the rows above to specific moments in the filtration timeline.
    }\label{fig:cubical_filtration}
\end{figure}

This procedure is visually demonstrated in Figure~\ref{fig:cubical_filtration} using a modified Mexican hat function as a toy model. As the threshold parameter increases from $T=-1.8$ to $T=2.8$, it acts analogously to a rising water level across the scalar field. Regions where the field's value falls below the threshold become ``flooded'' and enter the filtration as active blue cells within the discrete cubical complex. By computing the topological invariants at each step\,--- specifically the birth and death of isolated spatial components ($H_0$) and loops ($H_1$)\,--- the complex global geometry of the field is distilled into the persistence barcode shown in the bottom panel. In our case, we will utilize only $H_1$ homologies, as the number of $H_0$ is several orders higher, and it would take significantly longer computing time to compute later distances between the persistent diagrams. 

\section{21-cm line brightness field modeling}
\label{data}

We generate rectangular 21-cm line lightcones with \textsc{21cmFAST}
\citep[see, e.g.,]{mesinger2011}, a semi-numerical code that produces three-dimensional realizations of the redshifted 21-cm signal by combining the excursion-set formalism with perturbation theory. Starting from Gaussian initial conditions for a given cosmology, the density and velocity fields are evolved with Lagrangian perturbation theory, and the ionization field is obtained from the excursion-set prescription of \citet{furlanetto2004, mesinger2007, mesinger2011}: a cell is flagged as ionized when the cumulative number of ionizing photons exceeds the number of neutral hydrogen atoms, accounting for recombination, within a sphere of radius $R \le R_{\rm mfp}$. We compute the spin temperature self-consistently, including Ly$\alpha$ (Wouthuysen--Field) coupling and X-ray heating. When computing the persistent diagrams, we've used $\delta T_{21} = T_{21} - \bar{T_{21}}$ maps instead of a plane $T_{21}$ to account for the different mean effect values on diagrams.

The cosmological parameter sets are selected as follows. For each of the considered parameters (normalized Hubble parameter $h_0$, matter density parameter $\Omega_m$, matter fluctuations amplitude $\sigma_8$, and the spectral index of scalar matter perturbations $n_s$), we produce a simulation set in which it is the only changing parameter, keeping the others equal to the corresponding values in Planck cosmology (that is $\Omega_m=0.315$, $\sigma_8=0.811$,$h_0=0.67$, and $n_s=0.965$). In each set, there are 4 cases, which differ in equal steps in the selected parameter. The values are given in Table \ref{tab:1}.

To take into account cosmic variance for each case with a certain composition of cosmological parameters, we produce $[N_{\rm sim}]=7$ independent simulations with different initial conditions (different seeds) in a comoving box of $300\,\mathrm{Mpc}$ sampled on a $200^3$ grid, from which we extract brightness-temperature sub-volumes of $200\times200\times1$ voxels (two transverse directions and the line-of-sight/frequency axis) at redshifts spanning $z=7$--$30$ along the lightcone. Each field is passed through the identical topological pipeline of Section~\ref{cub}, so that the Gaussian mocks and the \textsc{21cmFAST} fields are treated on the same footing. The astrophysical parameters in all the simulations are the same; we set them equal to the fiducial ones given in Table~1 of \citep{park2019}.
\begin{wrapfigure}{r}{0.48\textwidth}
    \centering
    \includegraphics[width=0.48\textwidth]{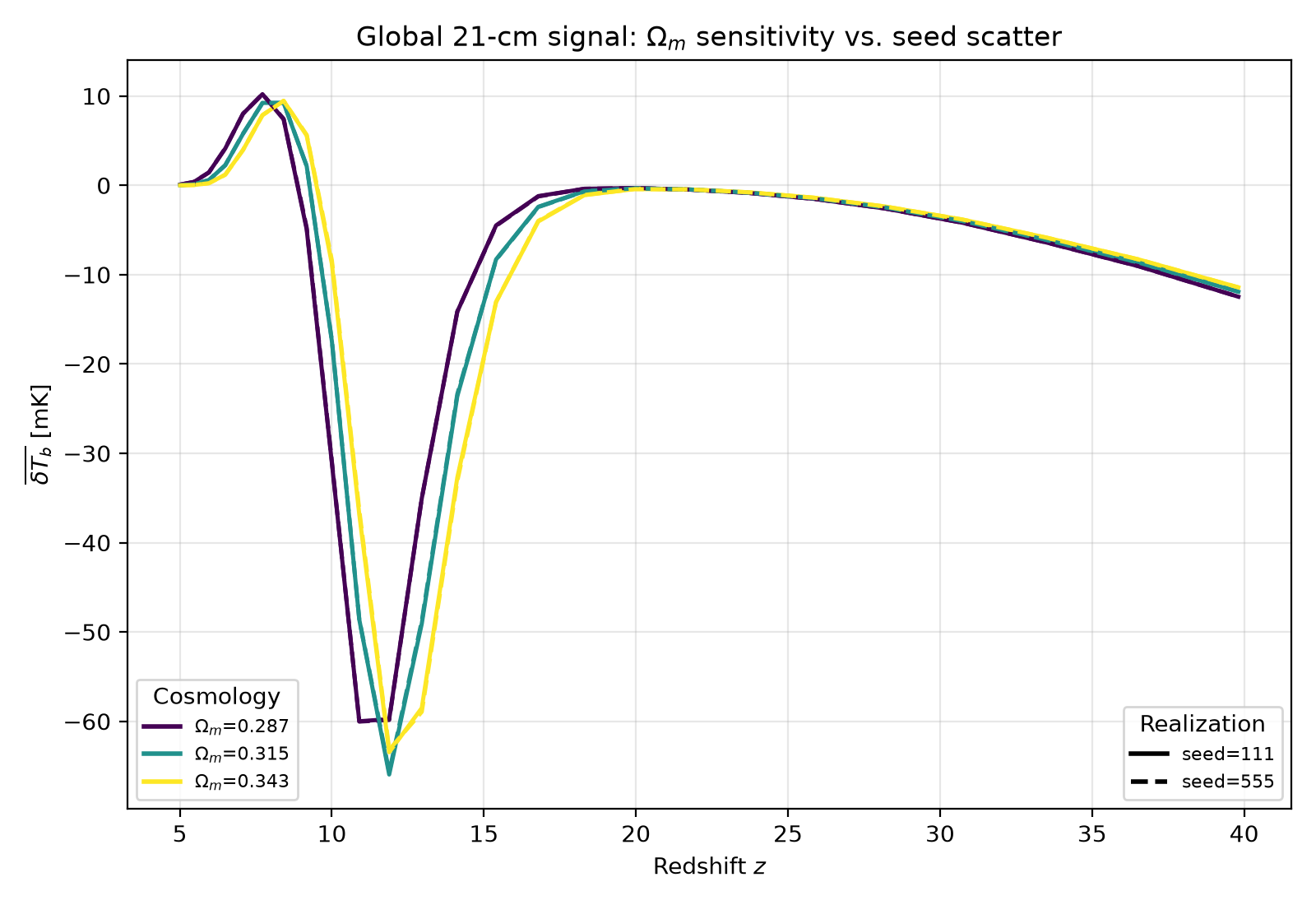}
    \caption{Globally averaged 21-cm brightness temperature as a function of redshift for $\Omega_m=0.287$, $0.315$, and $0.343$. For illustration,
we show results for two initial-condition seeds, $111$ and $555$.}
    \label{fig:glob}
\end{wrapfigure}

We analyze two-dimensional sky-plane slices rather than the full three-dimensional lightcone. Besides matching the per-channel form of interferometric images and providing redshift-resolved summaries, this choice avoids sensitivity to line-of-sight connectivity in the simulated lightcones. In 21cmFAST, lightcones are built by interpolating between coeval boxes evolved from a single set of initial conditions within a periodic volume, so structure along the line of sight is partly set by the node spacing and repeats on the scale of the box. A three-dimensional filtration would propagate these construction effects into the persistence diagrams.

\begin{figure}[htbp]
    \centering
    \includegraphics[width=0.45\textwidth]{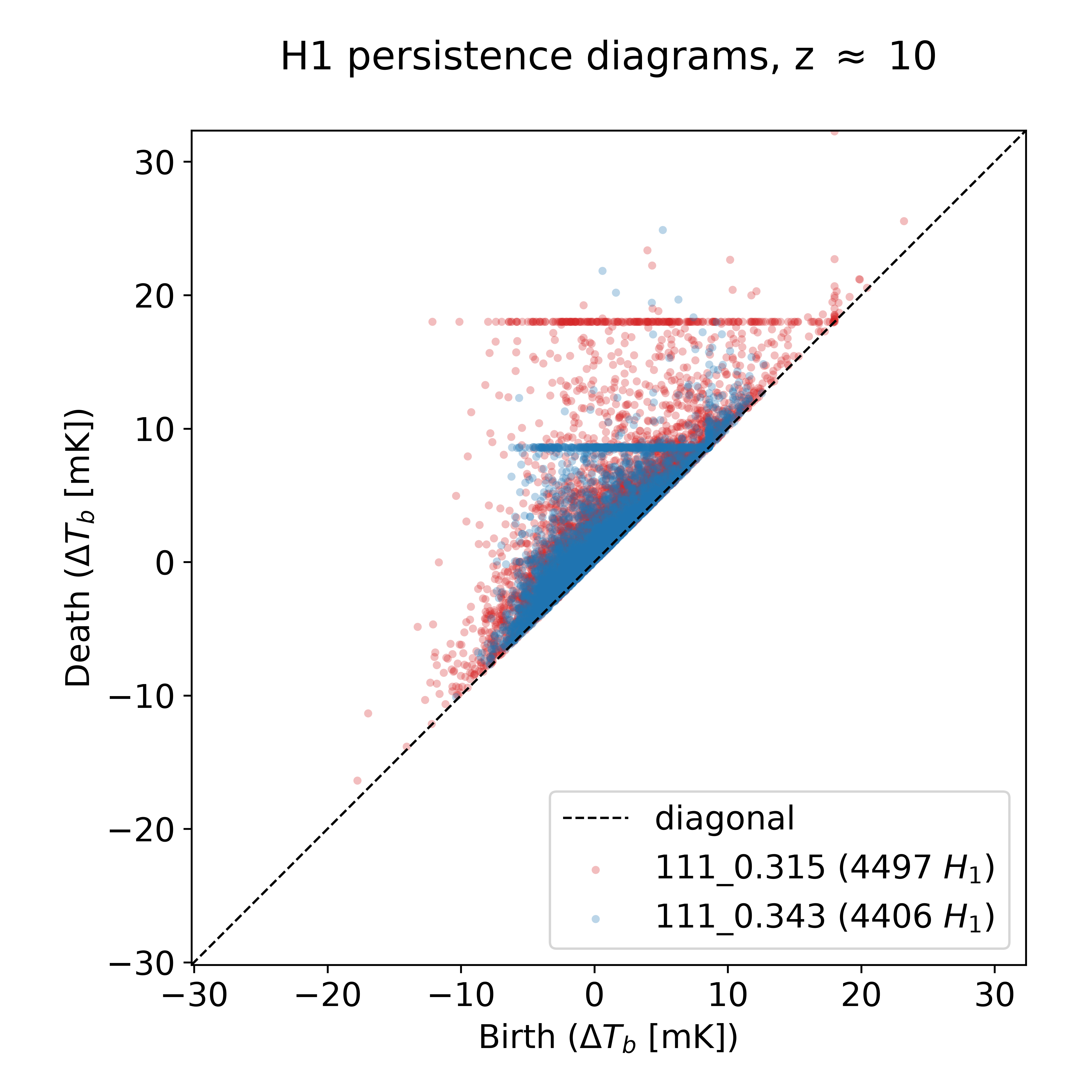}
   \includegraphics[width=0.45\textwidth]{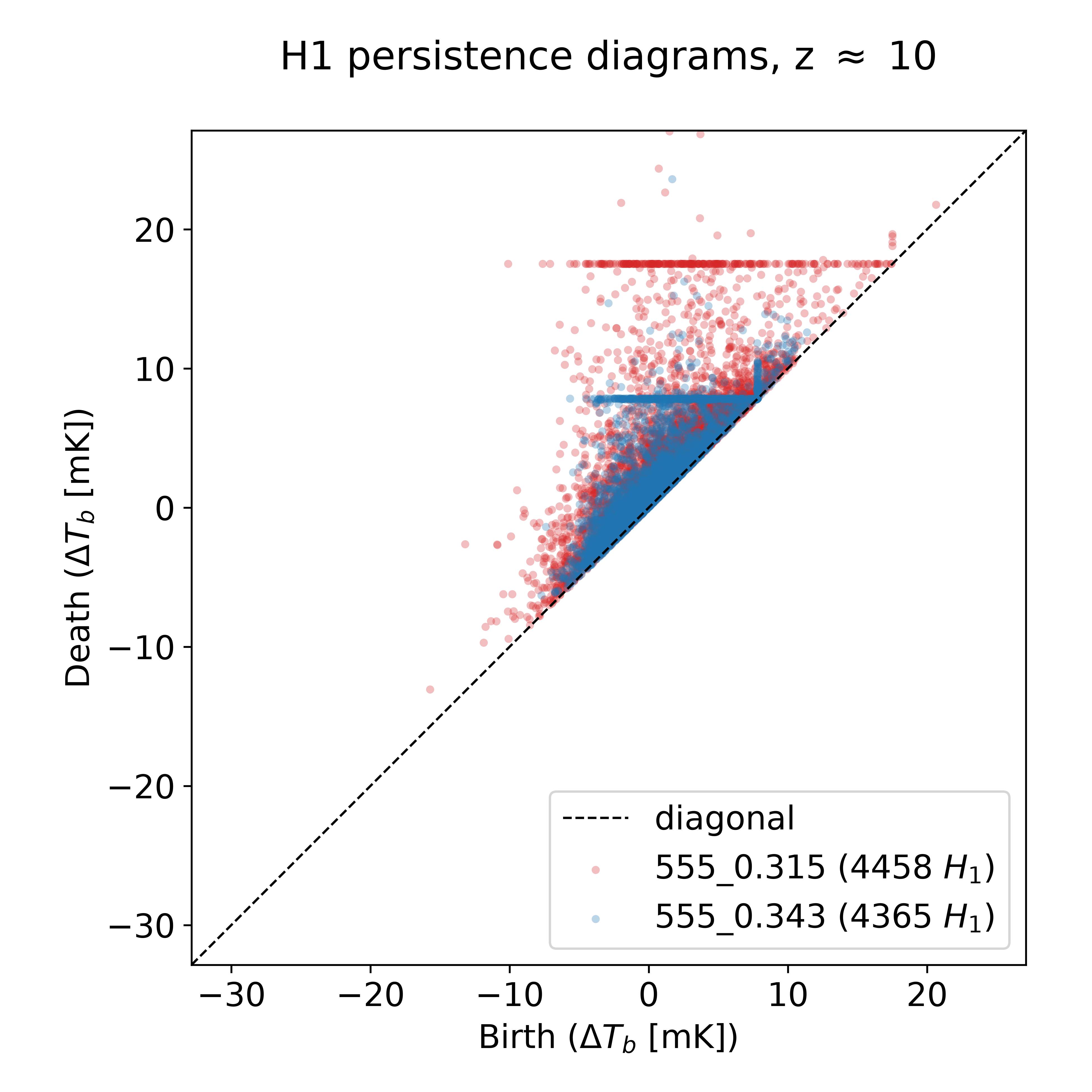}
    \caption{Examples of $H_1$ persistence diagrams at $z\simeq10$ for
$\Omega_m=0.315$ (red) and $\Omega_m=0.343$ (blue), using initial-condition
seeds $111$ (left) and $555$ (right). Each point represents a loop,
with coordinates indicating its birth and death brightness-temperature
thresholds above/below the mean value.}
    \label{fig:pd_comparison}
\end{figure}

In Fig. \ref{fig:pd_comparison}, examples of persistent diagrams are given. Visually, the homologies of simulations of different cosmologies are easily distinguishable.  We provide a global signal for different $\Omega_m$ values and two different seeds in Fig. \ref{fig:glob}. As can be seen from the figure, the average signal is negative in brightness temperature for the most interesting range of redshifts z $\in [9..11]$. This means that, on average, the 21-cm line absorbs the CMB light in this range. 

\section{Results}
\label{res}
Our results focus on measuring distances between persistent diagrams of 21-cm hydrogen line brightness fields generated with different cosmologies at a range of redshifts. To measure distances between the persistent diagrams of the 21-cm line field, we use the notion of the  Wasserstein distance. One can find a rigorous definition and a detailed explanation of the Wasserstein distance in \citep{tsizh2023}; below, we provide a brief overview.

Wasserstein distance (W-distance) is suitable for quantifying the similarity between two persistent diagrams of the same dimensionality, even with different numbers of homologies on each diagram. Its basic idea is to create many matches between dots on diagrams and then choose the one that best characterizes the difference between diagrams. This is achieved through the procedure of taking the infimum across the sums of Euclidean distances in all the pairs of given matching:
$$W_p(X, Y) = \left( \inf_{\eta} \sum_{x \in X} \|x - \eta(x)\|^p \right)^{1/p} $$

\begin{figure}[htbp]
    \centering
    \includegraphics[width=0.45\textwidth]{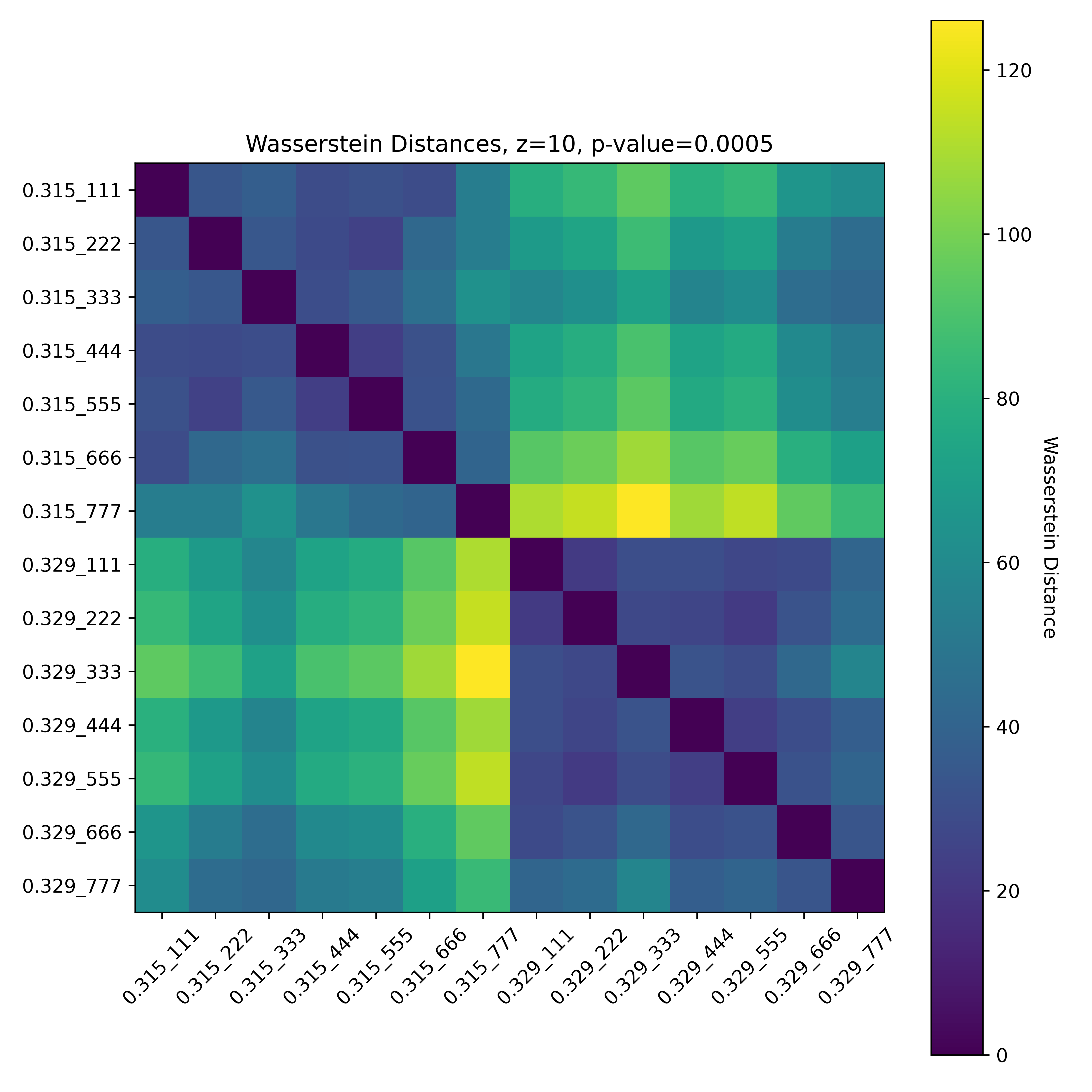}
    \includegraphics[width=0.45\textwidth]{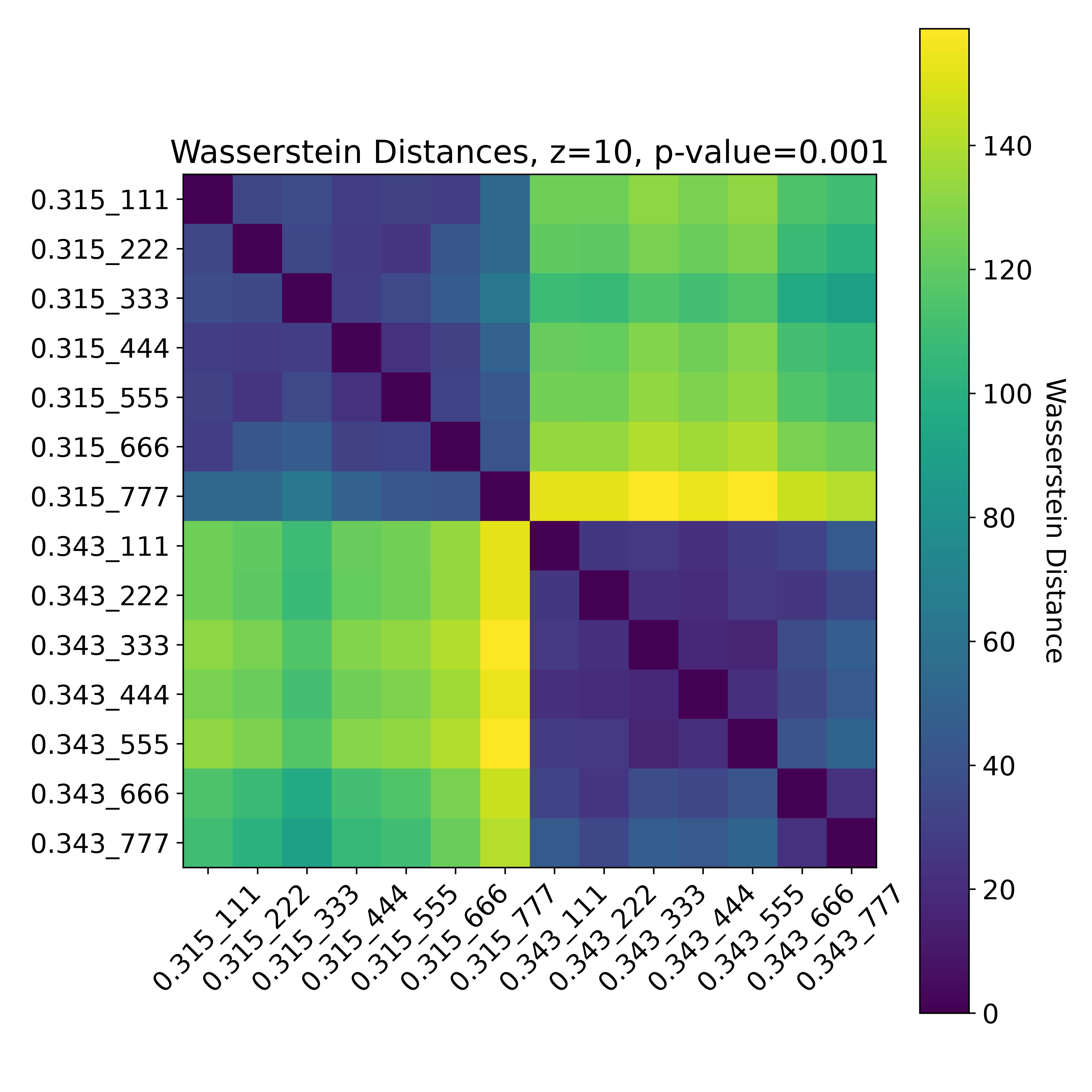}
    \caption{Examples of pairwise Wasserstein distance matrices between $H_1$ persistence diagrams at $z=10$, comparing $\Omega_m=0.315$ with $\Omega_m=0.329$ (left) and $\Omega_m=0.343$ (right). Each cosmology is represented by seven realizations corresponding to the initial-condition seeds ($111$--$777$). Distances between different cosmologies are generally larger than those within the same cosmology, with a stronger contrast in the right panel.}
    \label{fig:dist_mat}
\end{figure}
Here $W_p (X, Y)$ is the W-distance between two persistent diagrams, $x$ is one of the points on the first diagram, $\eta(x)$ is the corresponding matching of this point on the second diagram, $\|\cdot\|$ denotes the $L_\infty$ metric, and $p=2$. The sum is taken over all the points on the diagrams, and the infimum is taken over the set of matchings. The definition of W-distance is taken from \citep{Lacombe18}. For computing persistent diagrams and W-distances, we have used the package \texttt{Gudhi} in Python. All the scripts are available online to reproduce our results\footnote{Available on repo: \url{https://github.com/mtsizh/persistent-diagrams-distances-21cmline-field}}.

\subsection{Distinguishing the different cosmologies via clean 21-cm maps}
We build the W-distance matrices for subsets of simulations with different cosmologies and the same values of the initial seed generator. An example of such matrices is shown in Fig.~\ref{fig:dist_mat}. A satisfactory result one could expect from this build is that W-distances inside the group of simulations of the same cosmology are smaller, on average, than distances to simulations with one of the cosmological parameters changed.

Next, we use the PERMANOVA (Permutational analysis of variance) statistical test defined in \citep{Anderson2001} to determine the p-value of the hypothesis that different cosmologies can be distinguished with the W-distance built on their 21-cm line brightness field. PERMANOVA is a non-parametric statistical test designed to compare the distribution of groups of objects in some parameter space. For us, this parameter space is the space of relative W-distances. To be able to use the PERMANOVA test, we have to compare the subset of simulations of the same parameter pair-wise. Each subset contains 7 different seed simulations; hence, one would get a 14x14 W-distance matrix. We tested 2 cases for each cosmological parameter: small and large steps in the parameter. Such a pair of W-distance matrices is given in Fig. \ref{fig:dist_mat}. The small step roughly corresponds to a $2\sigma$ difference and the large step to a $2\times2\sigma$ difference. The details are given in Table \ref{tab:1}. 

\begin{figure}[htbp]
    \centering
    \includegraphics[width=0.45\textwidth]{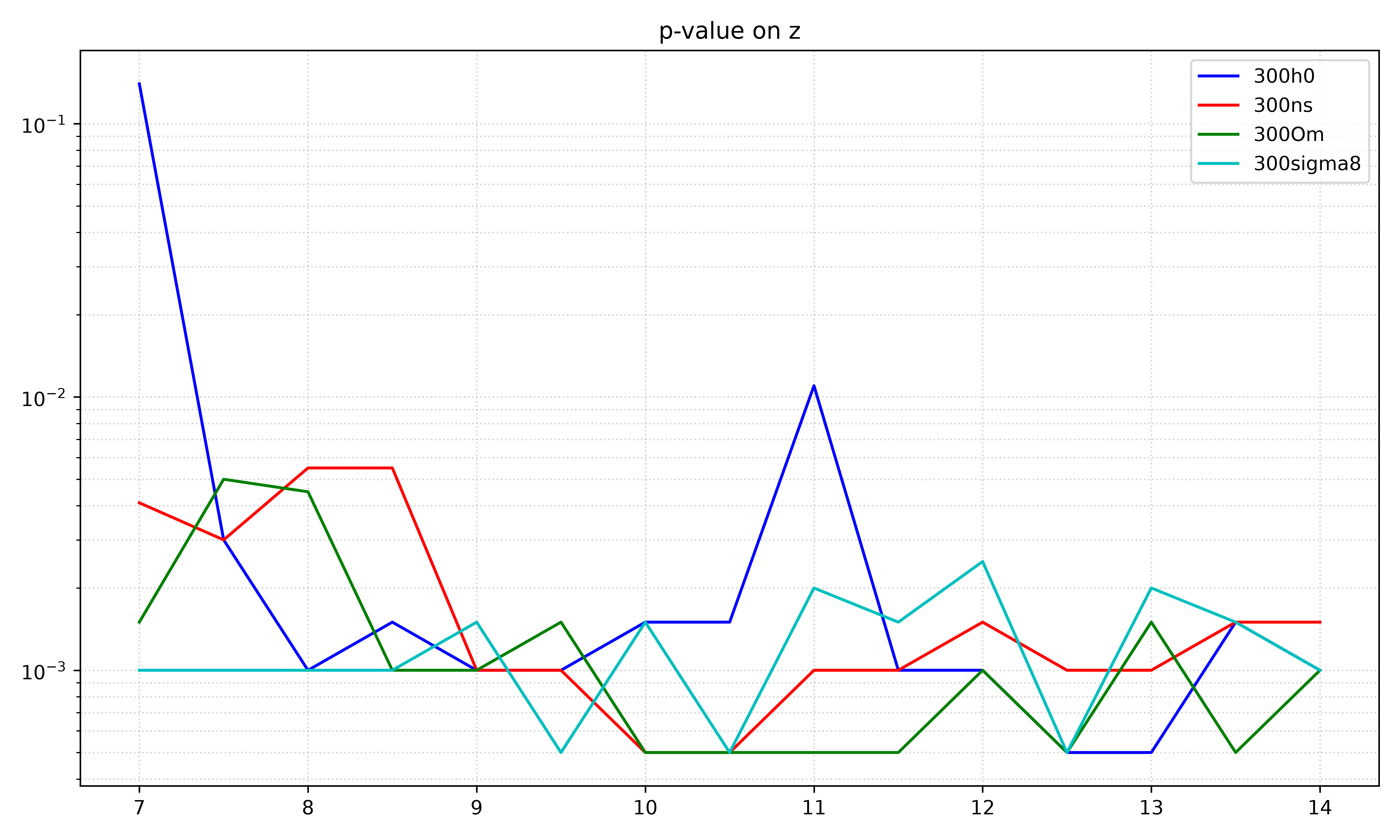}
    \includegraphics[width=0.45\textwidth]{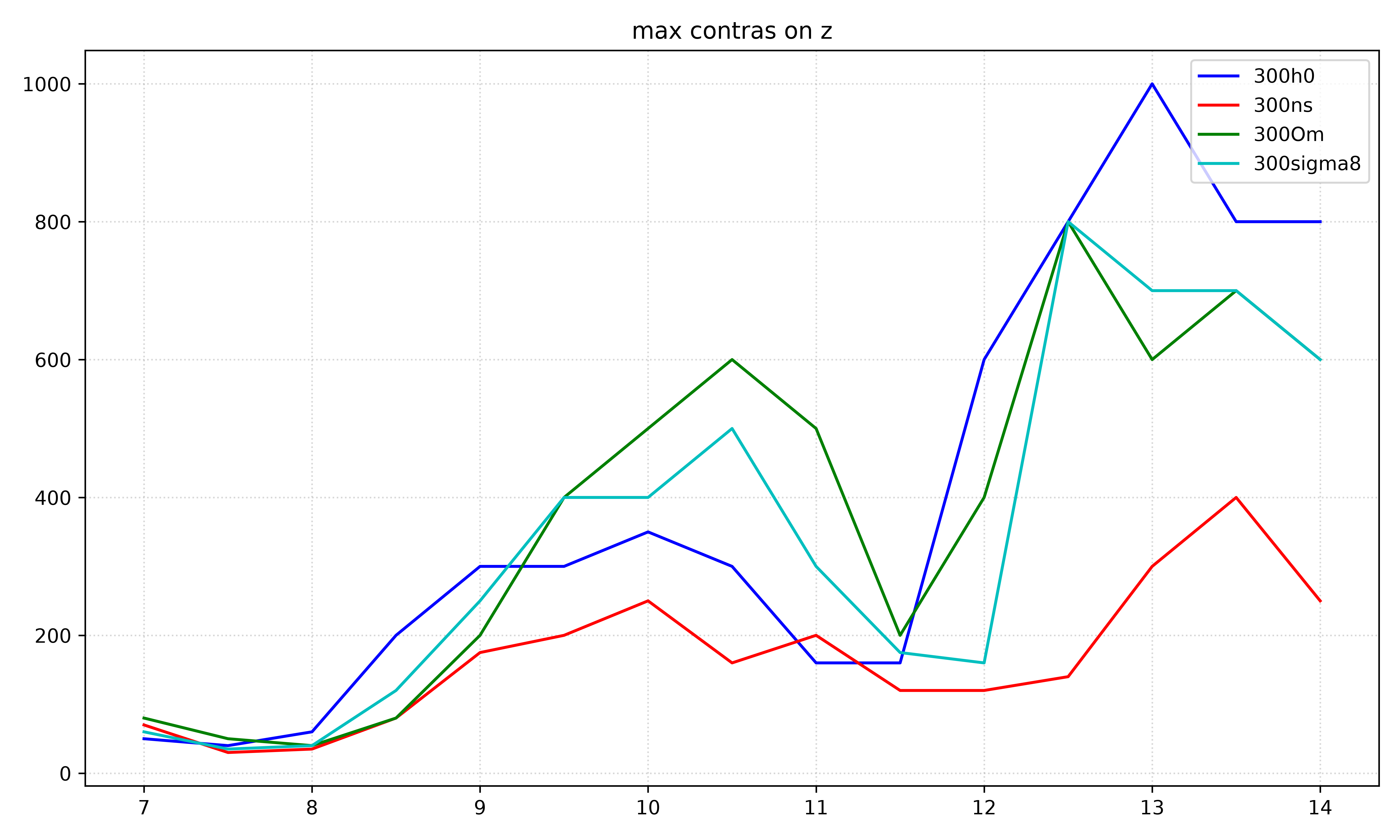}
    \includegraphics[width=0.45\textwidth]{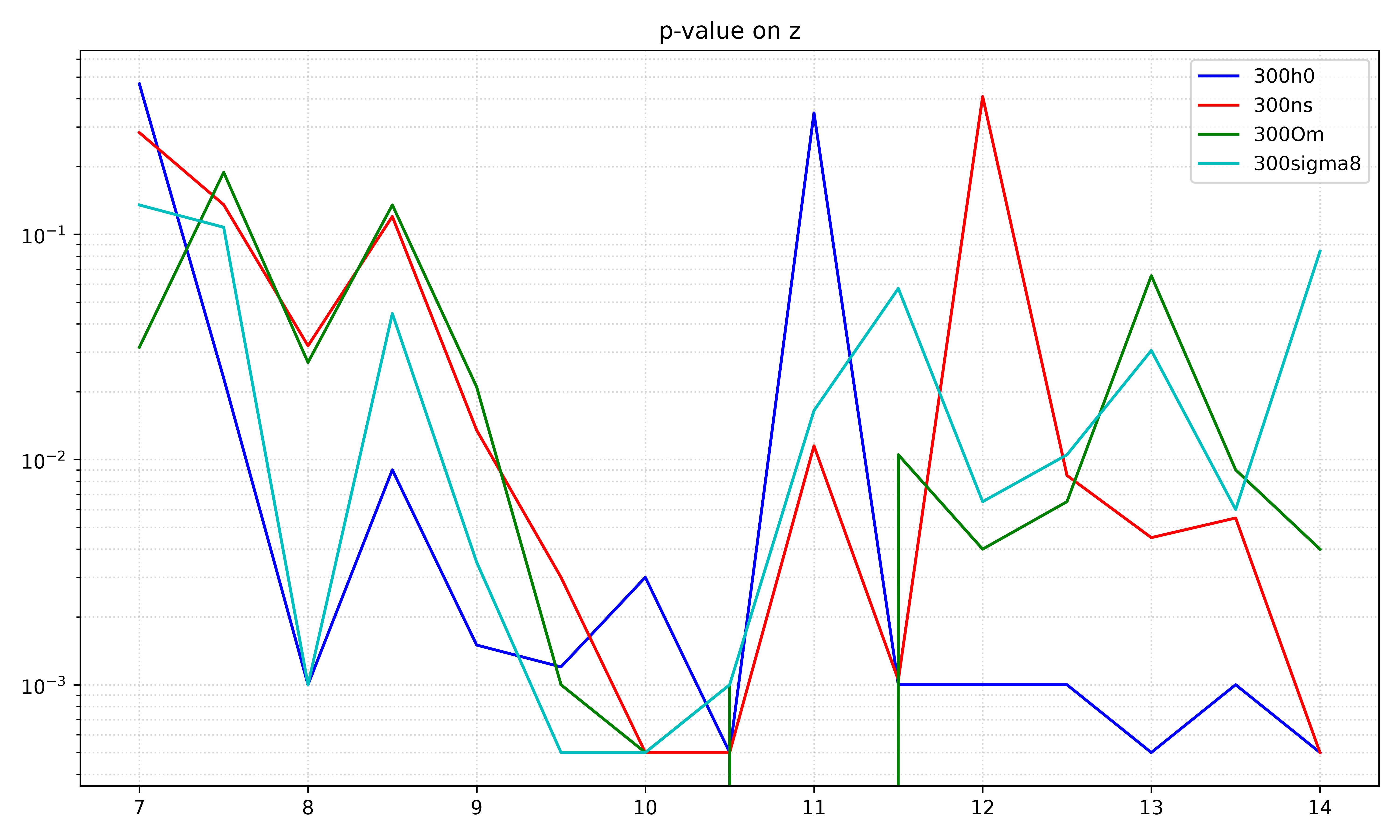}
    \includegraphics[width=0.45\textwidth]{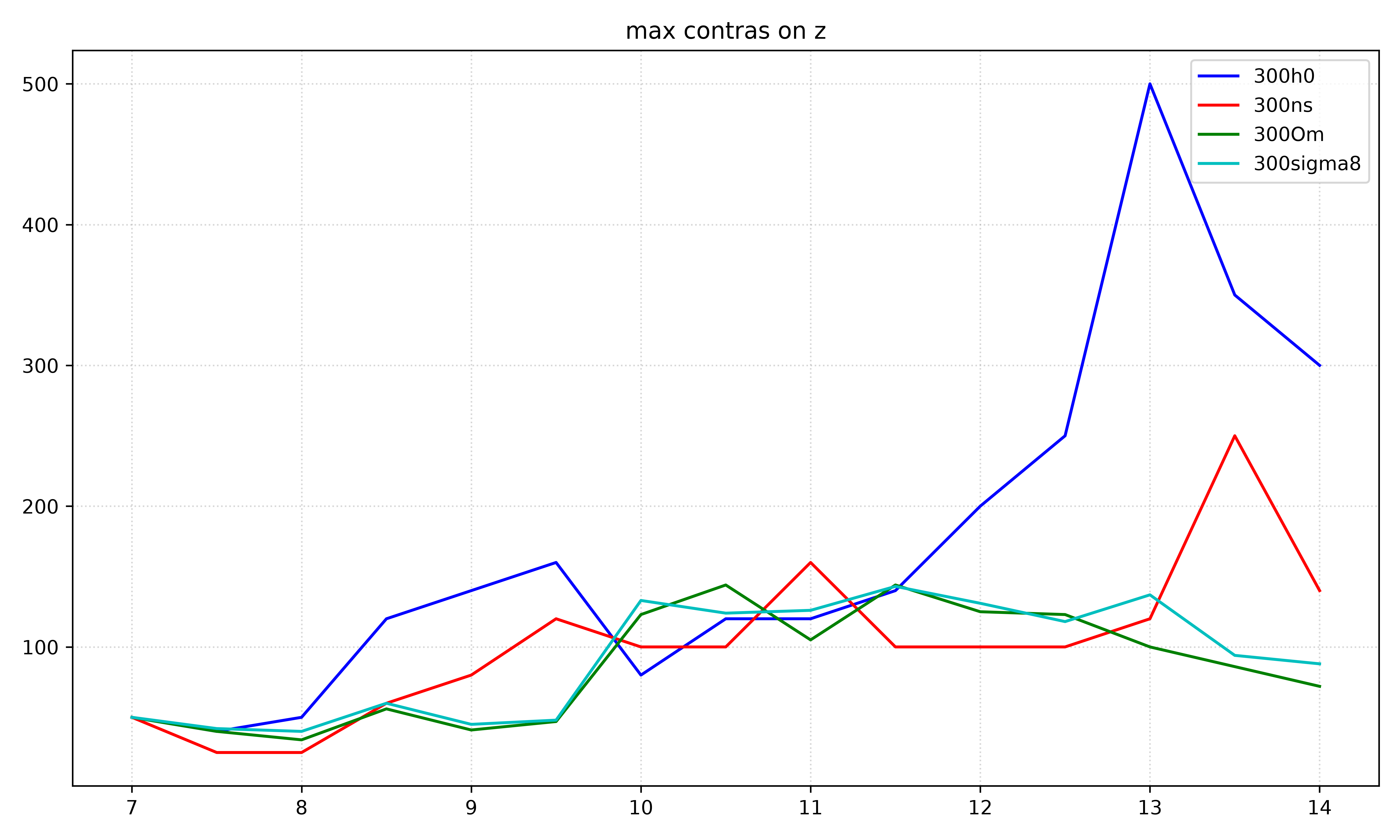}
    \caption{p-values for PERMANOVA test and maximum contrast in the W-distance matrices for a range of redshifts between 7 and 14. Different colors correspond to different cosmological parameter changes. Top panels: large step change. Bottom panels: small step changes, according to Table 1.}
    \label{fig:p_and_max}
\end{figure}

\begin{table}
    \centering
    \begin{tabular}{cccc}
         Cosmological parameter & Mean value & Small step &  Large step \\
         $h_0 $ & 0.67 & 0.02 & 0.05\\
         $n_s $ & 0.965 & 0.005 & 0.01\\
         $\Omega_m $ & 0.315 & 0.014 & 0.028 \\
         $\sigma_8 $ & 0.811 & 0.012 & 0.024\\
    \end{tabular}
    \caption{Cosmological parameters used in the simulations}
    \label{tab:1}
\end{table}
We present results for p-values and maximal values in the W-distance matrix for all the cosmological parameters for both small and large steps in the parameter difference in Fig. \ref{fig:p_and_max}.
Across the range of redshifts between 7 and 14 with a 0.5 step, we repeat this sequence to find the redshifts that will have the combination of the lowest p-values and highest contrast in the W-distance matrices. Thus, we will gain knowledge of which frequencies are the best for observing the 21-cm brightness map when one would like to determine the cosmological parameters with these observations. We can clearly see that the interval between redshifts 9.5 and 10.5 is the best for highlighting the difference between cosmologies for all parameters. We can also note that the p-value is slightly smaller for $n_s$ and $\Omega_m$ than for $h_0$ and $\sigma_8$; however, for all of them, p-values are well below 0.005 in the mentioned redshift interval. This is also reflected in terms of the highest W-distances - intervals with lower p-values correspond to the intervals with higher maximum W-distances.

\subsection{Distinguishing cosmological scenarios in mock SKA-Low observations}

In the previous subsection, we tested the ability of our approach to distinguish between cosmological models using idealized, noiseless 2D 21-cm maps. Realistic observations, however, are affected by foreground contamination, thermal noise, and the incomplete $uv$ coverage of the interferometer. We now assess how well our approach can distinguish between the cosmological scenarios under SKA-Low-like observing conditions.

We model SKA-Low thermal noise and instrumental sampling using
\texttt{tuesday}\footnote{\url{https://github.com/21cmfast/tuesday/tree/better_lc_noise}} and \texttt{21cmSense}\citep{Murray2024}, adopting the full AA4 configuration with baselines between $10\,\mathrm{m}$ and $2\,\mathrm{km}$. We assumed $6\,\mathrm{h}$ of observation per day over 1670 days, with $60\,\mathrm{s}$ integrations, a receiver temperature of $40\,\mathrm{K}$, and a sky temperature $T_{\rm sky}=60(\nu/300\,\mathrm{MHz})^{-2.55}\, \mathrm{K}$. We consider three treatments, all of which apply the $uv$ sampling to the 21-cm signal plus thermal noise: (i) no foreground-wedge removal, and (ii) ``optimistic'' and (iii) ``moderate'' foreground-wedge excision \citep{Pober:2014}.  Independent noise realizations were generated for each lightcone and target-redshift chunk, while identical underlying noise draws were used across the three treatments. Wedge excision was performed in three-dimensional Fourier space on chunks of 200 consecutive slices, spanning $300\,\mathrm{Mpc}$ in comoving distance and centered on each target slice, without a frequency taper. After filtering, we extracted the central two-dimensional map from each chunk and subtracted its spatial mean, similarly to the previous section.

\begin{figure}[htbp]
    \centering
    \includegraphics[width=1.05\textwidth]{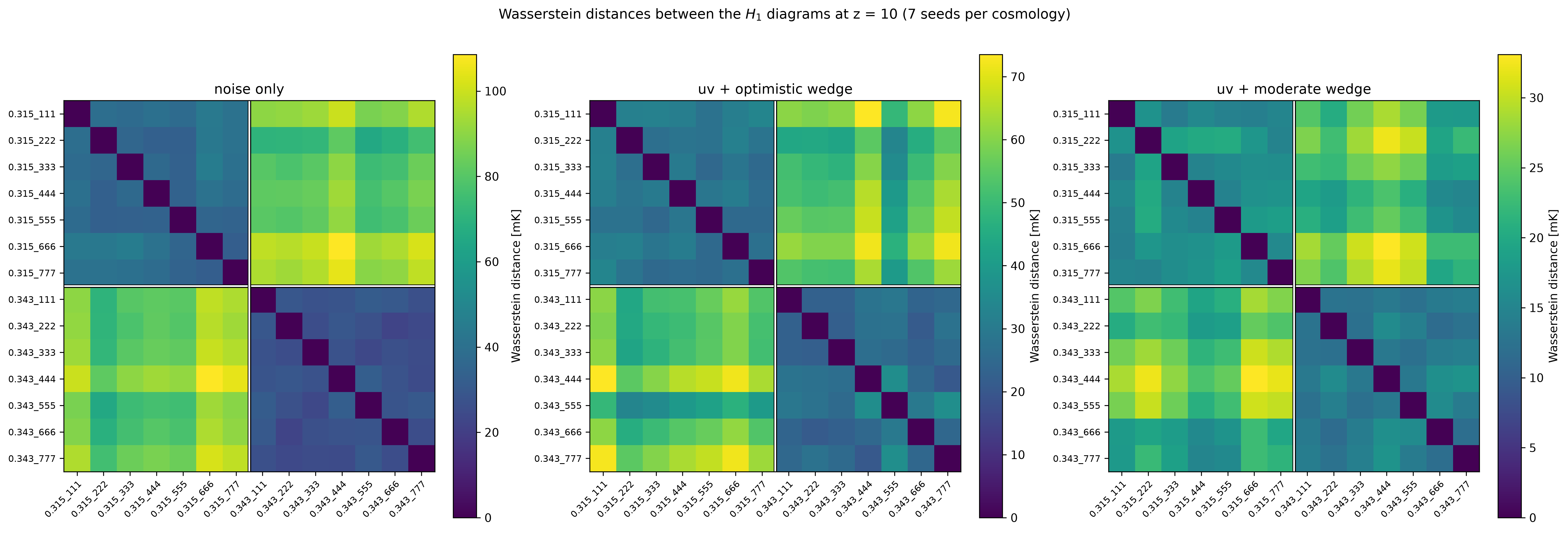}
    \caption{Examples of pairwise Wasserstein distance matrices between $H_1$ persistence
diagrams at $z=10$, comparing $\Omega_m=0.315$ with $\Omega_m=0.343$.
    \textit{Left:} noiseless slice plus SKA-Low thermal noise for a $10\,020$\,h
  observation, added in the measured $uv$ cells.
  \textit{Middle:} signal plus the same noise realization, sampled in the $uv$
  plane, with the optimistic foreground wedge removed in 3D on a $300$\, Mpc cubic
  chunk centered on the slice.
  \textit{Right:} as the middle panel, with the moderate wedge (horizon plus
  $0.1\,h\,{\rm Mpc}^{-1}$).
}
    \label{fig:noisy_w_m}
\end{figure}

For illustrative purposes, we compare the persistent homologies of the 2D fields corresponding to the cosmological models with $\Omega_m=0.315$ and $0.343$. The corresponding Wasserstein matrices for $z=10$ are shown in Fig.~\ref{fig:noisy_w_m}.

While noise and wedge removal make the two cosmologies harder to tell apart than in the noiseless case, they remain statistically distinguishable. For example, at $z=10$, the mean Wasserstein distance between slices from different cosmologies exceeds the mean distance between slices from the same cosmology by $8.6$, $5.8$ and $4.4$ times the scatter of the latter for treatments (i)--(iii), respectively. To test this more robustly, we apply a ``blocked'' PERMANOVA test, in which the cosmology labels are permuted only within each pair of lightcones that share a seed. With seven seeds, this gives $2^7/2 = 64$ distinct permutations, so the smallest attainable p-value is $1/64$. We obtain $p = 1/64$ at $z \simeq 10$--$11$ for treatments (ii) and (iii).

\section{Discussion \& Conclusions}
\label{concl}
In our work, we propose a new cosmological test by applying topological data analysis to simulated 2D maps of the 21-cm hydrogen line brightness temperature probed at some redshifts (within a single frequency channel). We have built the persistent diagrams with cubical complex filtration for a set of 21cmFAST cosmological simulations. We then computed the Wasserstein distance between the persistent diagrams for simulations with different cosmological parameters, $h_0$, $n_s$, $\Omega_m$, and $\sigma_8$, and different initial seeds. Using these distances as a differentiating value, we then performed a PERMANOVA statistical test to check the hypothesis that  W-distances can be used to tell apart the cosmologies that produced these 21-cm brightness maps. 

We searched for the redshift at which the difference between persistence diagrams caused by varying the cosmological parameters is largest relative to the scatter caused by varying the random initial conditions. We've found that this contrast peaks near $z \simeq 10$. In our fiducial model, this roughly corresponds to the moment when the global 21-cm signal vanishes because X-ray binaries have heated the IGM to a kinetic temperature close to that of the cosmic microwave background, $T_K \simeq T_{\rm CMB}$, and the spin temperature is coupled to the kinetic one. Regions heated by X-rays appear in emission and colder voids remain in absorption. We obtain a two-signed brightness field with a mean close to zero, and this field is expected to be non-Gaussian. We compute the topology of 2D 21-cm images with the mean subtracted, as for interferometric observations, which are insensitive to the global signal. Subtracting the mean only shifts all birth and death values in the persistence diagrams by a constant. The differences we find between cosmological scenarios therefore come from the morphology of the fluctuation field, not from differences in the global signal.

It is useful to place our results in the context of previous studies that have extracted astrophysical and cosmological information from 21-cm images, power spectra and topologies. Some studies have explored whether the physics of the epoch of reionization can be constrained using 2D 21-cm images at fixed redshifts, corresponding to observations in individual frequency channels. \citet{Hassan:20} applied convolutional neural networks to such maps and showed that cosmological parameters can, in principle, be inferred from these data. The neutral hydrogen fraction can also be estimated from individual 2D 21-cm images \citep[see, e.g.,][]{Mangena:20, Kwon:20}. These results are in line with our finding that even a single frequency channel carries information useful for discriminating between cosmological models.

Others have investigated cosmological inference from the 21-cm power spectrum. \citet{Schneider2023} performed an MCMC forecast analysis using a halo model of reionization to constrain cosmological parameters. \citet{Acharya2025} investigated cosmological models with different values of $\sigma_8$ and $h_0$ using LOFAR observations of the 21-cm power spectrum. At lower redshifts, $z<3$, \citet{Berti2023} used a multipole expansion of the 21-cm intensity-mapping power spectrum to forecast cosmological constraints from SKA observations. In all these cases, the limitations obtained from this analysis put worse limits on cosmological parameters, in comparison to our work, which confirms that TDA methods can gain additional information from data, in comparison to the power spectrum.

Our way of separating model differences from cosmic variance is closest to \citet{elbers2023}: they simulated several 21-cm lightcones of each model from different random initial conditions and compared the two groups with a permutation test, using the mean squared W-distance between persistence diagrams as the test statistic. 

Our work retains some systematic limitations. First of all, in this proof-of-concept work, we did not vary the astrophysical parameters and did not analyze the degeneracy between their cosmological parameter effects due to the high computational cost. It may be done within simulation-based inference applied to the pre-computed sample of the simulations. Secondly, we didn't include the mini-halos in the simulations for the same reason. More realistic analysis should obviously include them, as mini-halos are an important part of the ionization budget during the Epoch of Reionization. Another limitation is the number of initial seeds we simulated. We produced only 7 different seed simulations for each cosmology; increasing this number can probably yield even better p-values. The considered statistical test allows one to compare only two cosmologies at a time, in contrast to power spectra comparison. However, an advantage is also hidden here: we used only 2D slices of 21-cm line maps, which contain much less information than a 3D map from which spectra are built.

We have performed a test for the statistical distinguishability of selected cosmological models with TDA methods, rather than estimating parameters with posterior distributions, which is a more difficult problem. For both parameter separations considered, we find that the PERMANOVA test yields $p<0.005$ near $z\sim10$ for the clean 21-cm maps. Even after adding instrumental noise and removing modes within the foreground-contaminated wedge, the test yields $p\simeq0.02$ near this redshift for the two cosmologies with $\Omega_m=0.343$ and $0.315$, selected for this illustrative comparison. These results provide evidence that persistent-homology summaries are sensitive to variations in $h_0$, $n_s$, $\Omega_m$, and $\sigma_8$. Therefore, we argue that their further development as summary statistics for cosmological inference from the 21-cm brightness-temperature field is possible. Combining them with simulation-based inference is left for future work.

\section{Acknowledgments}
This work was done in the framework of the project “Tomography of the Dark Ages and Cosmic Dawn in the lines of hydrogen and the first molecules as a test of cosmological models” (state registration number 0124U004029), which was supported by the National Research Foundation of Ukraine. MT would also like to thank Prof. Andrei Mesinger, Dr. Jordan Flitter, and Dr. Nikolaos Triantafyllou for the fruitful discussion and technical help in producing the 21cmFAST simulations.

We used Claude Opus 5.5 (Anthropic) and ChatGPT Astra-6 (OpenAI) to improve the language and clarity of the manuscript, and Claude Code (Anthropic) to assist in writing the analysis scripts. The authors reviewed and verified all AI-assisted text and code and take full responsibility for the content of this publication.

\bibliographystyle{unsrtnat}
\bibliography{mybib}

\end{document}